\documentclass[sigconf]{acmart}
\usepackage{multirow}
\usepackage{tcolorbox}
\usepackage{listings}
\AtBeginDocument{%
  }

\setcopyright{acmlicensed}
\copyrightyear{2018}
\acmYear{2018}
\acmDOI{XXXXXXX.XXXXXXX}
\acmConference[Conference acronym 'XX]{Make sure to enter the correct
  conference title from your rights confirmation email}{June 03--05,
  2018}{Woodstock, NY}
\acmISBN{978-1-4503-XXXX-X/2018/06}

\begin{document}

\title{Enhancement Report Approval Prediction: A Comparative Study of Large Language Models}

\author{Haosheng Zuo}
\email{211250074@smail.nju.edu.cn}
\affiliation{%
  \institution{Nanjing University}
  \city{Nanjing}
  \country{China}
}

\author{Feifei Niu}
\email{feifeiniu96@gmail.com}
\affiliation{%
  \institution{University of Ottawa}
  \city{Ottawa}
  \country{Canada}
}

\author{Chuanyi Li*}
\email{lcy@nju.edu.cn}
\affiliation{%
  \institution{Nanjing University}
  \city{Nanjing}
  \country{China}
}

\begin{abstract}
Enhancement reports (ERs) serve as a critical communication channel between users and developers, capturing valuable suggestions for software improvement. However, manually processing these reports is resource-intensive, leading to delays and potential loss of valuable insights. To address this challenge, enhancement report approval prediction (ERAP) has emerged as a research focus, leveraging machine learning techniques to automate decision-making. While traditional approaches have employed feature-based classifiers and deep learning models, recent advancements in large language models (LLM) present new opportunities for enhancing prediction accuracy.

This study systematically evaluates 18 LLM variants (including BERT, RoBERTa, DeBERTa-v3, ELECTRA, and XLNet for encoder models; GPT-3.5-turbo, GPT-4o-mini, Llama 3.1 8B, Llama 3.1 8B Instruct and DeepSeek-V3 for decoder models) against traditional methods (CNN/LSTM-BERT/GloVe). Our experiments reveal two key insights: (1) Incorporating creator profiles increases unfine-tuned decoder-only models’ overall accuracy by 10.8\% though it may introduce bias; (2) LoRA fine-tuned Llama 3.1 8B Instruct further improve performance, reaching 79\% accuracy and significantly enhancing recall for approved reports (76.1\% vs. LSTM-GLOVE’s 64.1\%), outperforming traditional methods by 5\% under strict chronological evaluation and effectively addressing class imbalance issues. These findings establish LLM as a superior solution for ERAP, demonstrating their potential to streamline software maintenance workflows and improve decision-making in real-world development environments. We also investigated and summarized the ER cases where the large models underperformed, providing valuable directions for future research.
\end{abstract}





\maketitle

\section{Introduction}

Enhancement reports (ER) play a crucial role in software development, as they capture users' feedback and suggestions for improving applications. Consider a typical example from our dataset: ``Closing multiple windows warning should count tabs, too'' - a well-structured request that was ultimately approved. In contrast, vague proposals like ``blank page'' often face rejection due to unclear implementation value. Users may contribute enhancement or issue reports regardless of time and location, making feedback collection an ongoing and distributed process during software evolution. ER capture user expectations and serve as an essential means of engagement, offering developers insight into evolving user needs~\cite{feifeiniu2022}. Furthermore, as highlighted by the CHAOS Report~\cite{Hastie2015}, user involvement is one of the critical factors influencing project success, making ER a valuable tool for fostering user engagement and satisfaction. Unlike bug reports that address functional defects, ER focus on recommending new features or improvements that may not be immediately critical but can significantly enhance the competitiveness and usability of the software.

However, managing these reports poses significant challenges, particularly for large-scale software projects with frequent updates. Prior research suggests that less than 30\% of submitted ER receive approval, with approval decisions largely depending on the reports' cost-benefit tradeoff and alignment with project priorities~\cite{Nizamani2018}. The manual assessment of enhancement requests demands substantial effort, which may delay decisions or cause valuable proposals to be overlooked. For instance, according to Erne et al.~\cite{erne2020investigating} , the effectiveness of user-driven innovation diminishes without timely managerial responses.

Automating enhancement report approval prediction (ERAP) has thus emerged as a vital research direction, aiming to reduce handling costs, accelerate decision-making, and enhance the implementation of valuable user suggestions. Early approaches primarily relied on traditional machine learning models, such as Multinomial Naive Bayes (MNB)~\cite{Nizamani2018} and Support Vector Machines (SVM)~\cite{umer2019sentiment}~\cite{nafees2021machine}, utilizing textual features like TF-IDF representations~\cite{Nizamani2018}. Later advancements integrated sentiment analysis and contextual embeddings using deep learning models like CNN~\cite{arshad2021using} and BERT~\cite{niu2023utilizing}, improving prediction accuracy. Furthermore, recent studies emphasized the importance of incorporating metadata, such as creator profiles~\cite{niu2023utilizing}, which can significantly influence the likelihood of approval. Creators with higher domain expertise or frequent contributions are often more likely to produce actionable and valuable suggestions, as demonstrated by Heppler et al~\cite{heppler2016cares}. Despite these strides, challenges persist, particularly in striking a balance between precision and recall across different classes.

Recently, the emergence of large language models (LLM) has opened new possibilities for ERAP. These models, encompassing both encoder-only architectures like BERT~\cite{devlin2018bert} and decoder-only architectures like GPT~\cite{radford2018improving}, offer advanced capabilities in understanding and generating natural language. To assess the potential of these modern techniques, this study investigates the performance of encoder-only and decoder-only LLM in ERAP, alongside traditional methods, aiming to identify the most effective approaches for addressing the ERAP challenge.

\noindent\textbf{RQ1: How do encoder-only models perform on ERAP?}
Encoder-only models, represented by the BERT family~\cite{devlin2018bert}, are widely utilized in classification tasks for their ability to effectively capture bidirectional context.
In this study, we fine-tuned multiple encoder-only models, using a labeled dataset with a binary classification setup.

\noindent\textbf{RQ2: How do decoder-only models perform in ERAP?}
Decoder-only models, such as GPT, excel in generative tasks and have been increasingly applied to classification problems through prompting and fine-tuning.
In this RQ, we explore the performance of several decoder-only models, including GPT-3.5-turbo~\cite{openai_gpt35_turbo_doc}, GPT-4o-mini~\cite{openai_gpt4omini_doc},  DeepSeek-V3~\cite{liu2024deepseek}, Llama 3.1 8B~\cite{dubey2024llama} and Llama 3.1 8B Instruct~\cite{dubey2024llama}, under different prompting strategies, ranging from simple zero-shot setups to carefully engineered few-shot examples.
For Llama 3.1 8B and Llama 3.1 8B Instruct, we additionally applied LoRA (Low-Rank Adaptation) fine-tuning~\cite{hu2022lora}.

\noindent\textbf{RQ3: How do encoder-only and decoder-only models compare, and how do large models perform compared to traditional methods?}
In this RQ, we compare the performance of encoder-only and decoder-only models and evaluate their effectiveness against traditional methods.

\noindent\textbf{Contributions:}  
This paper makes the following key contributions:  
\begin{itemize}  
    \item This is the first work to empirically investigate the performance of LLM on the ERAP task, evaluating a total of 18 LLM.
    \item We assess the effectiveness of encoder-only models in this task by fine-tuning multiple variants from the BERT family, analyzing their classification performance.
    \item We investigate the capability of decoder-only models in ERAP, evaluating their performance under various prompting strategies and fine-tuning approaches.  
    \item We provide a comparative analysis between LLM and traditional machine learning methods, highlighting their respective strengths and limitations in handling ERAP.  
\end{itemize}  

The rest of this paper is organized as follows. Section~\ref{sec:relatedwork} provides background information on ERAP, ER analysis, and LLM. Section~\ref{sec:design} details our methodology. Section~\ref{sec:result} presents the experimental results, analyzing the data for each research question to highlight the strengths and limitations of LLM. Section~\ref{sec:threats} explains threats to validity. Section~\ref{sec:conclusion} concludes the paper and discusses feture work.

\section{Background and Related Work} \label{sec:relatedwork}
Traditionally, analyzing and predicting the approval of ER relies on manually crafted features or simpler machine learning techniques.
The advent of LLM, such as BERT and GPT, has brought transformative changes to NLP tasks, enabling the processing of unstructured textual data with greater context awareness and predictive capabilities.

\subsection{Enhancement Reports Approval Prediction}
ERAP seeks to identify enhancement requests unlikely to be approved, ideally before manual triage begins. Prior studies have primarily explored textual features and sentiment analysis for approval prediction, with various machine learning and deep learning approaches proposed to improve classification accuracy.

Nizamani et al.~\cite{Nizamani2018} developed a binary classification approach using Multinomial Naive Bayes (MNB), where textual ER were transformed into TF-IDF vectors representing term frequency distributions. Building upon this, Umer et al.~\cite{umer2019sentiment} enhanced the classification model by integrating sentiment analysis, hypothesizing that the emotional orientation of user requests could impact approval decisions. Using SENTI4SD, they extracted sentiment features and reported slight yet consistent gains in accuracy.

Further extending sentiment-aware approaches, Arshad et al.~\cite{arshad2021using, arshad2021deep} leveraged Word2Vec embeddings alongside deep learning classifiers to capture both syntactic and semantic structures in report texts. Their findings suggest that sentiment information provides additional context for distinguishing between approved and rejected reports. In parallel, Nafees et al.~\cite{nafees2021machine} proposed a Support Vector Machine (SVM)-based model to automate ER classification, highlighting the feasibility of machine learning techniques for approval prediction.

Unlike these traditional approaches that primarily relied on textual and sentiment-based features, Cheng et al.~\cite{cheng2021convolutional} explored deep learning architectures, employing a CNN-based model that integrated sentiment and deep feature representations to enhance classification accuracy. Meanwhile, Niu et al.~\cite{niu2022towards}~\cite{niu2023utilizing}introduced the notion of just-in-time ERAP, where approval likelihood is assessed as reports are submitted, and further expanded their work by incorporating metadata such as creator profiles to refine predictions.

\subsection{Enhancement Reports Analysis}
ER go through time consuming manual review and management. To address this challenge, researchers have developed various automated analysis methods aimed at improving the efficiency and accuracy of ER processing.

Shi et al.~\cite{shi2017understanding} applied fuzzy rules and linguistic analysis to extract structured information from ER, categorizing sentences into predefined types such as intent, explanation, and benefit. This structured approach facilitates more systematic analysis of report content.
Jayatilleke et al.~\cite{Jayatilleke2018amethod} introduced a framework for requirements change analysis, identifying dependency relationships and estimating change complexity through a matrix-based approach.
Similarly, Morales-Ramirez et al.~\cite{morales2017exploiting} explored properties of user feedback to support requirements prioritization, proposing an automated ranking approach that aids in decision-making by identifying the most critical enhancement requests.

Beyond prioritization, classification methods have been proposed to further enhance the organization and retrieval of ER. Li et al.~\cite{li2018automatically} built a classification framework that leverages machine learning to automatically assign enhancement reports to categories like capability, usability, and security, thereby improving report organization and retrieval.

In addition to classification, code localization plays a crucial role in bridging enhancement reoports with relevant source code. Zhang et al.s~\cite{zhang2019where2change} proposed a cosine similarity-based technique with adaptive weighting to locate source code fragments closely linked to the described enhancements.
Furthermore, Xu et al.s~\cite{xu2018mulapi} introduced the MULAPI method, which recommends API methods for new feature requests by leveraging similarities between historical feature requests and API documentation. This approach aids in software evolution by streamlining API selection for new feature.

Overall, ER analysis encompasses multiple aspects, including structuring report content, prioritization, classification, code localization, and API recommendation.

\subsection{Large Language Models}

The evolution of LLM represents a paradigm shift in natural language processing, driven by architectural innovations, scaling laws, and novel training methodologies. Early language models, based on recurrent neural networks (RNNs) and long short-term memory (LSTM) architectures~\cite{hochreiter1997long}, struggled with long-range dependencies and computational inefficiency. The introduction of the Transformer architecture by Vaswani et al.~\cite{vaswani2017attention} revolutionized the field through its self-attention mechanism, enabling parallel computation and superior modeling of contextual relationships across arbitrary token distances.

The subsequent decade witnessed two dominant architectural paradigms: encoder-only and decoder-only models. BERT (Bidirectional Encoder Representations from Transformers)~\cite{devlin2018bert} pioneered the encoder-only approach by pretraining on masked language modeling (MLM) and next sentence prediction (NSP), enabling deep bidirectional context understanding. This architecture excelled in classification and extraction tasks, inspiring variants like RoBERTa~\cite{liu2019roberta}, which optimized training dynamics through larger batches and dynamic masking, and ELECTRA~\cite{clark2020electra}, which introduced a more sample-efficient replaced token detection objective. DeBERTa~\cite{he2021debertav3} further advanced encoder architectures by disentangling content and positional embeddings, enhancing semantic granularity.

Decoder-only models form the backbone of modern generative language models, employing an autoregressive training paradigm where each token is predicted sequentially based on preceding tokens. This approach enables high-quality text generation and has been central to the development of models like the GPT series~\cite{radford2018improving, radford2019language, brown2020language, achiam2023gpt}.
A fundamental breakthrough in the evolution of these models is the discovery of Scaling Laws~\cite{kaplan2020scaling}, which establish a predictable relationship between model performance, parameter count, dataset size, and computational budget. Research has shown that larger models, when trained on sufficiently diverse and high-quality data, exhibit emergent abilities that smaller models lack, such as few-shot and zero-shot learning. This insight guided the development of models like GPT-3, which, with 175 billion parameters, demonstrated the ability to generalize across a wide range of tasks with minimal task-specific training~\cite{brown2020language}. Later iterations, such as GPT-4~\cite{achiam2023gpt}, refined these capabilities by incorporating multimodal reasoning and improved alignment techniques.
Beyond proprietary models, open research efforts have contributed significantly to the advancement and democratization of LLM. Projects such as Llama~\cite{touvron2023llama}~\cite{touvron2023llama2}~\cite{dubey2024llama} have challenged the notion that performance is solely dictated by model size, showing that architectural optimizations and training efficiency can yield competitive results even with fewer parameters. Meanwhile, DeepSeek-V3~\cite{liu2024deepseek}, employing Mixture-of-Experts (MoE) architectures, achieves superior efficiency by dynamically activating only a subset of parameters during inference, significantly reducing computational overhead.

Despite the impressive performance of large-scale decoder-only models, their resource demands remain a critical limitation. To enable broader adoption and fine-tuning for specialized tasks, parameter-efficient adaptation techniques have gained traction. Among those, Low-Rank Adaptation (LoRA) has emerged as a effective approach~\cite{hu2022lora}. Instead of updating the full model, LoRA injects trainable low-rank matrices into frozen pre-trained weights, significantly reducing memory consumption while maintaining fine-tuning flexibility. This method has proven valuable for domain-specific applications where training a full-scale model from scratch is infeasible.

\section{Study Design}\label{sec:design}

We outline our experimental design for addressing the ERAP task using both encoder-only and decoder-only models.

\subsection{Dataset Preparation}

The dataset used in this study was constructed by replicating the data collection and preprocessing methodology detailed by Niu et al. \cite{niu2023utilizing}. Following their approach, we utilized Bugzilla's REST API to retrieve ER and extracted metadata fields including \texttt{id}, \texttt{summary}, \texttt{type}, \texttt{resolution}, \texttt{creator}, \texttt{creation\_time}, \texttt{assigned\_to}, etc. The detailed meaning of each metadata field is provided in Table~\ref{tab:metadata}.

\begin{table}
  \caption{Metadata of Enhancement Reports (reproduced from Niu et al.~\cite{niu2023utilizing})}
  \label{tab:metadata}
  \begin{tabular}{lp{0.65\columnwidth}}
    \toprule
    \textbf{Metadata} & \textbf{Description} \\
    \midrule
    \texttt{id} & The unique numeric ID of this report. \\
    \texttt{product} & The name of the product this report is in. \\
    \texttt{severity} & The current severity of the report. \\
    \texttt{resolution} & The current resolution of the report (empty if it is open). \\
    \texttt{type} & The type of it, including enhancement, defect, or task. \\
    \texttt{priority} & The priority of the report. \\
    \texttt{summary} & The summary of this report. \\
    \texttt{description} & The actual text of the report. \\
    \texttt{creator} & The name of the person who filed this report (the reporter). \\
    \texttt{creation\_time} & When the report was created. \\
    \texttt{assigned\_to} & The name of the user to whom the report is assigned. \\
    \bottomrule
  \end{tabular}
\end{table}

Bugzilla categorizes issue reports into three types: enhancement, defect, and task. In line with Niu et al., we focused exclusively on enhancement reports. Report descriptions were obtained from the first comment submitted by the creator, using the corresponding report ID. We collected reports from the ten most active applications on Bugzilla, spanning submissions from September 10, 1997, to July 13, 2016.

The \texttt{RESOLUTION} metadata field indicates whether a report has been resolved. However, reports with a \texttt{NULL} resolution status were excluded from the dataset, as they do not provide clear information about their resolution status. The final dataset consists of 40,551 ER across 10 applications.

The resolution status in the \texttt{RESOLUTION} field includes the following categories: \texttt{INVALID}, \texttt{DUPLICATE}, \texttt{FIXED}, \texttt{WONTFIX}, \texttt{INCOMPLETE}, \texttt{WORKSFORME}, \texttt{EXPIRED}, \texttt{MOVED}, and \texttt{INACTIVE}. Among these, only \texttt{FIXED} indicates a successfully resolved report. To ensure comparability with other state-of-the-art approaches, we adopted their dataset partitioning strategy, labeling \texttt{FIXED} reports as \textit{APPROVED} and all others as \textit{REJECTED}. The distribution of resolution statuses across different applications is shown in Table~\ref{tab:resolution_status}.

\begin{table}
  \caption{Distribution of Resolution Status in Different Applications (reproduced from Niu et al.~\cite{niu2023utilizing})}
  \label{tab:resolution_status}
  \begin{tabular}{lccc}
    \toprule
    \textbf{Application} & \textbf{APPROVED} & \textbf{REJECTED} & \textbf{TOTAL} \\
    \midrule
    Bugzilla & 2176 & 2422 & 4598 \\
    SeaMonkey & 924 & 7111 & 8035 \\
    Core Graveyard & 367 & 1202 & 1569 \\
    Core & 2997 & 4790 & 7787 \\
    MailNews Core & 407 & 1771 & 2178 \\
    Toolkit & 422 & 1535 & 1597 \\
    Firefox & 842 & 6595 & 7337 \\
    Thunderbird & 451 & 3853 & 4304 \\
    Calendar & 461 & 1139 & 1600 \\
    Camino Graveyard & 347 & 839 & 1186 \\
    \midrule
    Total & 9394 & 31157 & 40551 \\
    \bottomrule
  \end{tabular}
\end{table}

Niu et al.~\cite{niu2023utilizing} demonstrated that incorporating the creator profile improves prediction accuracy. 
Following their approach, we constructed a creator profile for each ER creator. 
The creator profile (CP) consists of three key attributes: creator ID, role, and submission frequency. The creator ID is the creator's email address. The role attribute indicates whether the creator is a regular user, cross-application developer, or inner-application developer. 
We determined the creator’s role by analyzing the relationship between the \texttt{creator}, \texttt{assigned\_to} and \texttt{product} metadata fields. The submission frequency represents the total number of ER submitted by the creator across all applications.

To prevent the model from encountering data during training that is later than the data seen during testing, thereby avoiding data leakage, we applied a temporal split strategy. For each \texttt{product} separately, ER were sorted in ascending order based on \texttt{creation\_time}. Within each product, the earliest 80$\%$ of the reports were assigned to the training set, the next 10$\%$ to the validation set, and the final 10$\%$ to the test set. This ensures that the dataset split follows a strict chronological order within each product, making the evaluation more realistic and aligned with real-world scenarios.

\subsection{Large Language Models}
The selection of language models in this study is driven by their proven effectiveness across diverse NLP tasks.
We prioritize models that have demonstrated state-of-the-art performance in semantic understanding (encoder-based architectures) and contextual generation (decoder-based architectures). These models span both encoder-only and decoder-only architectures, leveraging their unique strengths for task-specific optimization. The models investigated in this work include:

\textbf{BERT-uncased (Base and Large)}~\cite{devlin2018bert} is a case-insensitive bidirectional transformer-based language model designed for deep contextual understanding of text. It employs a masked language model (MLM) pre-training objective, allowing it to capture both left and right context simultaneously. Additionally, BERT introduces a next sentence prediction (NSP) task to enhance sentence-level understanding. The model is pre-trained on massive corpora, including BooksCorpus and English Wikipedia, and achieves state-of-the-art performance on multiple NLP benchmarks. 
    
\textbf{RoBERTa (Base and Large)}~\cite{liu2019roberta} is a robustly optimized BERT model that improves upon the original BERT architecture by training with larger mini-batches over more data, removing the next sentence prediction objective, training on longer sequences, and dynamically changing the masking pattern applied to training data. 
    
\textbf{ELECTRA-discriminator (Base and Large)}~\cite{clark2020electra} is a transformer-based model that introduces a novel pre-training approach where a discriminator learns to distinguish between real tokens and those replaced by a generator.
    
\textbf{DeBERTa-v3 (Base and Large)}~\cite{he2021debertav3} is an enhanced transformer-based language model that improves upon its predecessor by incorporating ELECTRA-style pre-training with a replaced token detection objective and introducing gradient-disentangled embedding sharing to decouple the gradients between the embedding layer and the rest of the network.
    
\textbf{XLNet-cased (Base and Large)}~\cite{yang2019xlnet} is a transformer-based language model that leverages a novel permutation-based objective to capture bidirectional context without relying on masked token prediction. It integrates autoregressive and autoencoding approaches to model token dependencies more effectively.

\textbf{Llama 3.1 8B}~\cite{dubey2024llama} is a decoder-only language model from the Llama series, designed for high-quality text generation. It refines the original Llama architecture by incorporating advanced training optimizations and enhanced pre-training techniques to boost both performance and computational efficiency.
    
\textbf{Llama 3.1 8B Instruct}~\cite{dubey2024llama} is an instruction-tuned variant of the base Llama 3.1 8B model, refined through supervised fine-tuning on a diverse set of instruction-response pairs to better align with user commands.
    
\textbf{GPT-3.5-Turbo}~\cite{openai_gpt35_turbo_doc} is a variant of OpenAI's GPT-3.5 series, designed for conversational AI applications, offering improved performance in generating human-like responses.
    
\textbf{GPT-4o-mini}~\cite{openai_gpt4omini_doc, openai2024gpt4omini} is a lightweight variant of OpenAI’s GPT-4o, designed for cost-efficient and scalable deployment. It retains the core architecture of GPT-4o while being optimized for reduced computational requirements.
    
\textbf{DeepSeek-V3}~\cite{liu2024deepseek} is a Mixture-of-Experts (MoE) language model with 671B total parameters, of which 37B are activated per token. It features Multi-head Latent Attention (MLA) and DeepSeekMoE architectures for efficient inference and training. The model pioneers an auxiliary-loss-free strategy for load balancing and sets a multi-token prediction training objective for stronger performance.

To summarize, the models used in this study can be categorized into encoder-only and decoder-only architectures. Encoder-only models (e.g., BERT-base-uncased, BERT-large-uncased, RoBERTa-base, RoBERTa-large, DeBERTa-v3-base, DeBERTa-v3-large, ELECTRA-base-discriminator, ELECTRA-large-discriminator, XLNet-base-cased and XLNet-large-cased) are fine-tuned directly for classification, while decoder-only models (e.g., Llama 3.1 8B, Llama 3.1 8B Instruct, GPT-4o-mini, GPT-3.5-turbo and DeepSeek-V3) are optimized using advanced prompt engineering and LoRA fine-tuning techniques.

\subsection{Strategies for Encoder-only models}

\paragraph{Fine-Tuning Encoder-Only Models:} 
Encoder-only models, such as those in the BERT family, were directly fine-tuned for the binary classification task.
We fine-tuned mainstream decoder-only models on the labeled dataset for binary classification tasks.
The training process focused exclusively on textual features, ensuring the models effectively differentiated between positive and negative examples.
Standard fine-tuning was employed to optimize performance and adapt the models to the task.
Additionally, experiments were conducted with different learning rates to identify the most effective configuration for improving classification accuracy.
The specific training parameters for each model are provided in the code repository.

\subsection{Strategies for Decoder-only models}

We explored the performance of decoder-only models in ERAP tasks by leveraging various strategies, ranging from prompt-based methods to fine-tuning approaches. These strategies are designed to optimize decoder-only models for classification tasks, traditionally dominated by encoder-based methods. Below, we describe the key strategies employed:

\paragraph{Zero-shot Prompting:}
We formulated a structured prompt to directly query decoder-only models without providing explicit examples. Initially, we used only the input text of the ER as the basis for classification. Subsequently, we experimented with adding additional information, including profile of the creator and a detailed explanation of their roles. Different prompt commands and sequences were tested, including directly instructing the model to perform binary classification, as well as first prompting the model for a more detailed multi-class prediction before deriving the binary classification from its output. Excluding the binary fine-tuned Llama models, all other models exclusively used the latter approach.

\paragraph{Few-shot prompting with task-relevant examples:}
To enhance the performance of decoder-only models, we adopted a few-shot prompting strategy by incorporating relevant examples into the prompt. These examples were dynamically selected to ensure contextual relevance and diversity. Specifically, we first filtered examples by selecting those associated with the same product and submitted earlier than the input instance. We then computed semantic similarity using all-roberta-large-v1 sentence embeddings and selected the five most similar approved examples and the five most similar rejected examples. Each selected example included summary, description, creator profile, and its resolution label.

\begin{figure}[h]
    \centering
    \includegraphics[width=1\linewidth]{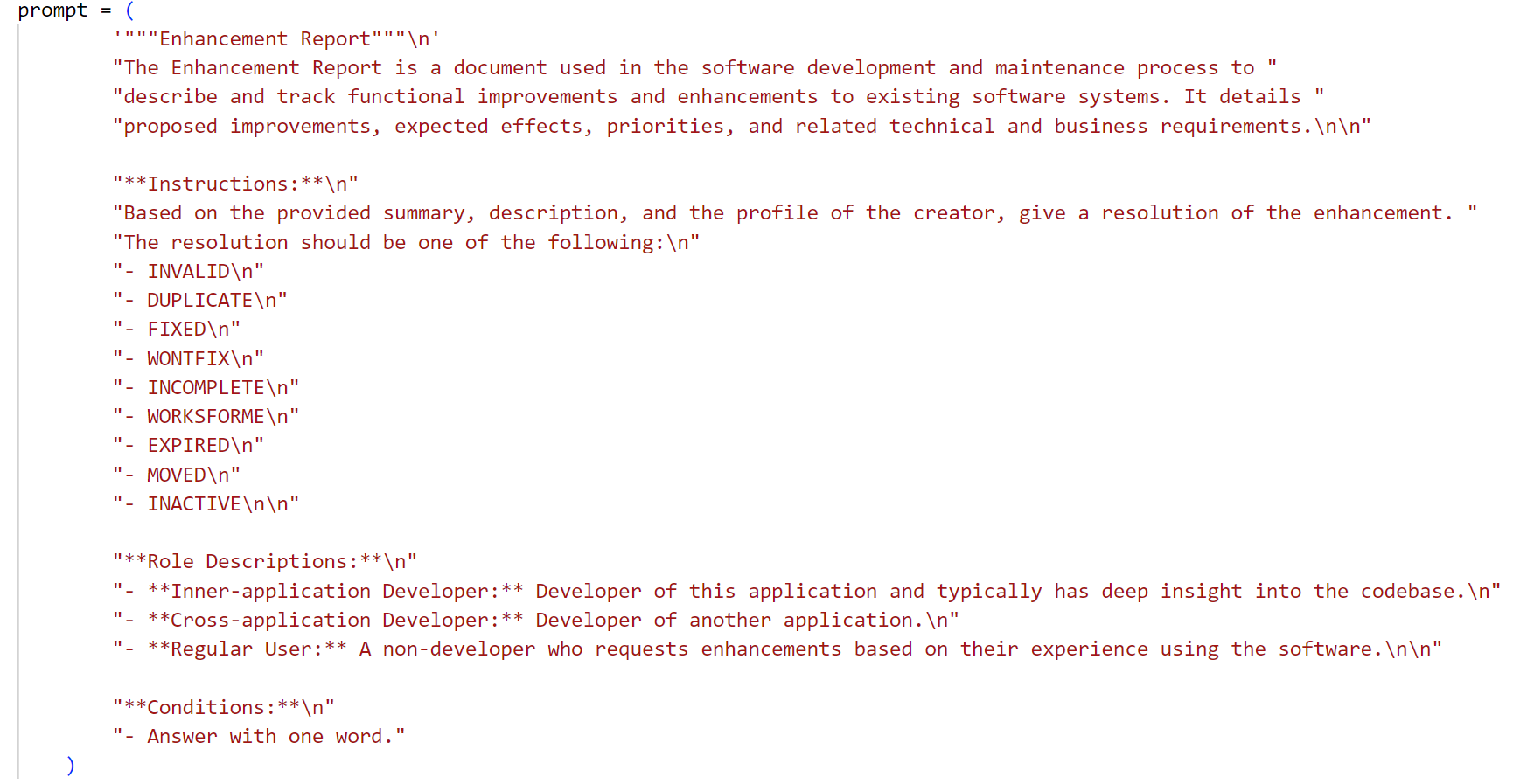} 
    \caption{Prompt}
    \label{fig:prompt}
\end{figure}

\paragraph{Structured prompt optimization:}
To maximize the effect, we iteratively refined the prompt structure. The optimized prompt incorporated the following key elements: (1) a clear and concise task description, (2) an explicit explanation of the creators' roles, (3) task-relevant examples categorized by resolution labels, and (4) systematic exploration of example placement, including positioning at the beginning, middle, and end of the prompt.
\noindent The final optimized prompt structure is illustrated in Figure~\ref{fig:prompt}.

The few shot examples are added in front of the prompt, creator profile and text (summary and description) are added after the prompt in that order. For direct binary categorization, only the tags FIXED and WONTFIX are retained in the prompt. In text-only input scenarios, the phrase ``and the profile of the creator'' is also removed from the prompt, ensuring that the model makes its prediction based solely on the summary and description.

\paragraph{LoRA-based fine-tuning:}
We fine-tuned the Llama 3.1 8B and Llama 3.1 8B Instruct models using two configurations tailored to the ERAP task. These configurations included: (1) training the model for multi-class classification followed by binary classification and (2) directly fine-tuning the model for binary classification, both incorporating creator profile during training. For each configuration, we tested model performance with and without examples. We applied the LoRA technique with a rank of 8, using a learning rate of 1e-4, a batch size of 1 per device, and gradient accumulation steps of 8 for an effective batch size of 8. Training was conducted for 3 epochs with FP16 precision, utilizing cosine learning rate scheduling with a warmup ratio of 0.1. The model was evaluated every 500 steps during training. Each LoRA fine-tuning process took approximately 16-18 hours on a single NVIDIA V100 GPU. These fine-tuning processes were facilitated using LLaMA-Factory~\cite{zheng2024llamafactory}.

\subsection{Baselines}
To ensure fair comparison with the most relevant prior work, we selected these representative methods from prior work on ERAP, rather than conducting an evaluation across all possible traditional machine learning classifiers. The state-of-the-art approach from the study of Niu et al. ~\cite{niu2023utilizing} employs a CNN-BERT hybrid model with creator profile integration, which serves as our primary baseline. Their results demonstrate that both CNN and LSTM classifiers achieve strong performance in ERAP, while BERT and GloVe embeddings effectively capture semantic information. Motivated by these findings, we adopt both CNN and LSTM as our classification models and incorporate BERT and GloVe for text representation to ensure a comprehensive evaluation. 

Following the methodology outlined in the study of Niu et al.~\cite{niu2023utilizing}, we reproduce these four approaches on our chronologically organized dataset to ensure a fair comparison.
\subsection{Evaluation Methodology and Metrics}

We evaluate the performance of our classifiers using accuracy, precision, recall  and F1-score. $TP$ represents the number of correctly predicted approved reports, $FP$ represents the number of rejected reports incorrectly predicted as approved, $FN$ represents the number of approved reports incorrectly predicted as rejected, and $TN$ represents the number of correctly predicted rejected reports. 

\begin{equation}
\text{Accuracy} = \frac{TP + TN}{TP + TN + FP + FN}
\end{equation}

\begin{equation}
\text{Precision} = \frac{TP}{TP + FP}
\end{equation}

\begin{equation}
\text{Recall} = \frac{TP}{TP + FN}
\end{equation}

\begin{equation}
F1 = \frac{2 \times \text{Precision} \times \text{Recall}}{\text{Precision} + \text{Recall}}
\end{equation}

\subsection{Implementation and Environment}

Except for deepseek-v3, all the open-source pre-trained models (i.e.,Llama 3.1 8B, Llama 3.1 8B Instruct, BERT-base-uncased, BERT-large-uncased, RoBERTa-base, RoBERTa-large, DeBERTa-v3-base, DeBERTa-v3-large, ELECTRA-base-discriminator, ELECTRA-large-discriminator, XLNet-base-cased and XLNet-large-cased) were downloaded from HuggingFace. 
For local execution and fine-tuning, Llama models were run with FP16 precision. Additionally, for all decoder-only models, the temperature was set to 0.

The experiments were carried out on a server running CentOS 8, equipped with an Intel Xeon Platinum 8268 CPU (20 cores allocated), four NVIDIA V100 GPUs (32GB each) and 220GB RAM. During training, only a single GPU was utilized at a time to manage resource allocation and simplify experimental control.

\section{Experimental Results}\label{sec:result}
\subsection{RQ1: Performance of Eecoder-only Models}

\begin{table*}[h]
    \caption{Experimental Results on Encoder-only models}
    \label{tab:table3}
    \centering
    \resizebox{\textwidth}{!}{
        \begin{tabular}{cccccccc|ccccccc} 
            \hline  
            \multirow{3}{*}{} & \multicolumn{7}{c|}{Text} 
                              & \multicolumn{7}{c}{Text+Creator Profile} \\ 
            \cline{2-15}  
                              & \multirow{2}{*}{Acc (\%)} 
                              & \multicolumn{3}{c}{Approved} & \multicolumn{3}{c|}{Rejected} 
                              & \multirow{2}{*}{Acc (\%)} 
                              & \multicolumn{3}{c}{Approved} & \multicolumn{3}{c}{Rejected} \\ 
            \cline{3-8} \cline{10-15} 
                              &                                           
                              & P (\%) & R (\%) & F (\%) & P (\%) & R (\%) & F (\%)
                              &                                           
                             & P (\%) & R (\%) & F (\%) & P (\%) & R (\%) & F (\%) \\ 
            \hline  
            bert-base-uncased & 77.6 & 78.4 & 63.7 & 70.3 & 77.1 & 87.5 & \textbf{82}   
                              & 74.6 & 78.6 & 53.7 & 63.8 & 73   & 89.6 & 80.5 \\
            bert-large-uncased & 76.1 & 76.1 & 62.2 & 68.4 & 76.1 & 86   & 80.8 
                               & 75.1 & \textbf{80.3} & 53.3 & 64.1 & 73.1 & 90.6 & 80.9 \\
            roberta-base & 78   & 74.1 & \textbf{72.6} & \textbf{73.4} & \textbf{80.7} & 81.9 & 81.3 
                         & 77.6 & 79   & 63.1 & 70.2 & 77   & 88   & 82.1 \\
            roberta-large & \textbf{78.4} & 75.9 & 70.7 & 73.2 & 80   & 84   & \textbf{82}   
                          & 74.3 & 76.8 & 54.8 & 63.9 & 73.2 & 88.2 & 80   \\
            electra-base-discriminator & 77.4  & 77.4 & 64.8 & 70.5 & 77.5 & 86.5 & 81.7 
                                       & \textbf{78.2}  & 78.9 & 65.1 & 71.4 & 77.9 & 87.6 & \textbf{82.4} \\
            electra-large-discriminator & 77.6 & 78.6 & 63.3 & 70.1 & 77   & 87.7 & \textbf{82} 
                                        & 76.7  & 79.5 & 59.5 & 68   & 75.5 & 89   & 81.7\\
            deberta-v3-base & 77.7  & 75.1 & 69.6 & 72.2 & 79.4 & 83.5 & 81.4 
                            & 77.6  & 77.5 & 65.3 & 70.9 & 77.7 & 86.5 & 81.9 \\
            deberta-v3-large & 77.2 & 77.1 & 64.4 & 70.2 & 77.3 & 86.3 & 81.5 
                             & 70.4 & 79.9 & 38.7 & 52.2 & 68   & \textbf{93.1} & 78.6 \\
            xlnet-base-cased & 77.2 & 74.1 & 69.5 & 71.7 & 79.2 & 82.6 & 80.9 
                             & 77.6 & 78.2 & 64.3 & 70.5 & 77.4 & 87.2 & 82 \\
            xlnet-large-cased & 74.8   & \textbf{78.9} & 53.8 & 64   & 73.1 & \textbf{89.8} & 80.6 
                              & 78.1 & 76.7 & \textbf{68.2} & \textbf{72.2} & \textbf{78.9} & 85.2 & 82 \\ \hline
        \end{tabular}
    }
    
\end{table*}
Table~\ref{tab:table3} shows encoder-only models achieving strong classification accuracy on ERAP, ranging from 70.4\% to 78.4\%. Among them, RoBERTa-large achieves the highest accuracy (78.4\%) on text-only inputs, while ELECTRA-base-discriminator performs best (78.2\%) when creator profile information is included. These results highlight the robustness of encoder-only architectures in ERAP tasks, with certain models benefiting from additional contextual signals.

From a class-wise perspective, the results show that most models achieve higher precision for the ``Approved'' class than recall, indicating a tendency to be more conservative in predicting approval. This reduces false positives but may lead to missing some valid enhancement requests. In contrast, recall values for the ``Rejected'' class are generally higher than those for the ``Approved'' class, often exceeding 80\%, indicating that models are more confident in correctly identifying rejected ERs.

Adding creator profile information impacts models differently. While models like RoBERTa-base and DeBERTa-v3-base maintain stable performance, others, such as XLNet-large-cased, see significant improvements. For instance, XLNet-large-cased shows a notable increase in recall for the ``Approved'' class (from 53.8\% to 68.2\%) and overall accuracy (78.1\%), indicating its ability to leverage additional contextual signals. However, DeBERTa-v3-large experiences a drop in accuracy (from 77.2\% to 70.4\%), suggesting that profile information may introduce noise for certain models.  Notably, the performance difference between different scale's encoder models was relatively small, potentially suggesting that for this specific task and dataset, the capabilities of smaller models are already approaching saturation.

\begin{figure}[h]
    \centering
    \includegraphics[width=1\linewidth]{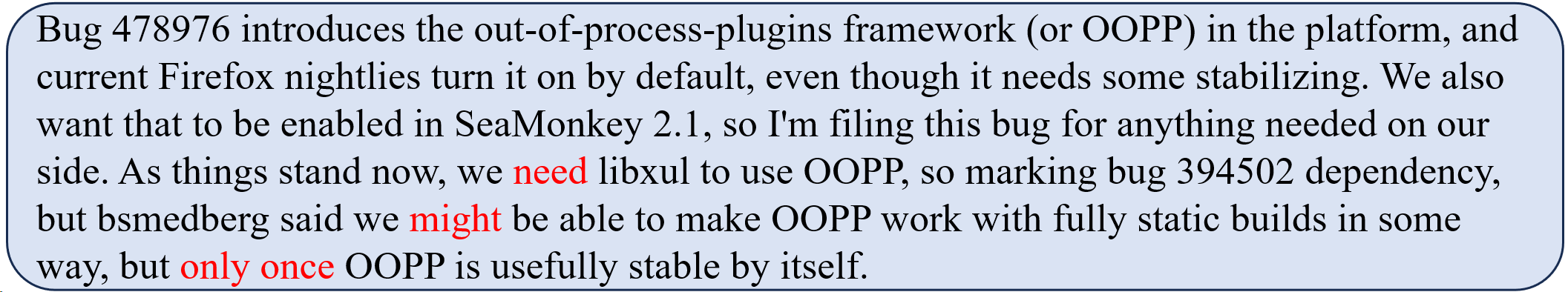} 
    \caption{Description}
    \label{fig:description}
\end{figure}

Interestingly, while the encoder-only models demonstrate high consistency in their misclassification patterns, XLNet-large-cased exhibits some distinct characteristics. Specifically, it rejects certain ER that all other models classify correctly. These rejected reports often involve system architecture changes or modifications affecting operational workflows. For example, the enhancement request ``Use out-of-process-plugins (OOPP) framework in SeaMonkey'' was rejected by XLNet-large-cased despite being approved by all other models. The report describes the integration of the OOPP framework into SeaMonkey 2.1, highlighting dependencies on existing bugs and stability concerns, as illustrated in Figure~\ref{fig:description}.

A possible explanation for this discrepancy is that the other models recognize the specificity and detailed justification provided in the enhancement request. The description explicitly outlines the current implementation status, dependencies, and stability considerations, making the enhancement request appear well-reasoned and feasible. Additionally, the careful wording in these descriptions, with phrases such as ``might'', ``only once'', and ``need'' introduces utious and conditional language. These terms suggest a measured approach rather than a firm demand for change, which may make the request more acceptable to these models.
In contrast, XLNet-large-cased may focus more on the uncertainty and dependencies mentioned in the description, interpreting them as signals that the enhancement is not yet ready for approval. Instead of viewing conditional phrasing as a sign of careful planning, it may assign greater weight to the underlying uncertainties, leading to stricter criteria when evaluating changes related to system architecture. This conservative tendency is reflected in the model's error patterns. Among the 1,280 enhancement requests where ten encoder-only models disagreed, XLNet-large-cased had the highest error rate at 48.83\%, slightly above BERT-large-uncased (44.61\%). However, its false rejection rate was particularly high, incorrectly rejecting 506 requests (68.56\%), far exceeding BERT-large-uncased (49.32\%), which had the second-highest false rejection count. Meanwhile, its false acceptance rate was notably low, misclassifying only 119 requests (21.96\%), the lowest among the models, even below ELECTRA-discriminator-large (30.81\%). This pattern is also observed in other enhancement requests, such as ``Set up Mac OS X 10.6 64-bit SeaMonkey (2.1) builders,'' where XLNet-large-cased rejects an ER that all other models accept. This suggests that XLNet-large-cased places more emphasis on stability concerns over forward-planning considerations, making it more conservative in accepting modifications that involve significant dependencies or potential risks.


Similarly, XLNet-large-cased exhibits frequent misclassification when dealing with ER related to UI modifications. Among the 1,280 disputed ERs, 610 are related to UI changes. Although the overall error rates of XLNet-large-cased do not differ significantly between UI-related and non-UI-related ERs, its false rejection rate for UI-related ER reaches 78.14\%, which is substantially higher than the general level. In contrast, the false acceptance rate is only 14.86\%. For example, the request ``Bugzilla login field should use 'placeholder' HTML5 attribute instead of JavaScript'' was incorrectly rejected by XLNet-large-cased and four other models. Additionally, ``Add UI to select maildir for storage when creating accounts'' was also incorrectly rejected by XLNet-large-cased and four partially overlapping models. These cases indicate that certain encoder-only models struggle with UI-related enhancement requests, potentially due to the lack of explicit architectural or functional reasoning within the textual descriptions. UI changes are often described concisely without detailed technical dependencies, which may lead to misinterpretations by models that prioritize dependency resolution and implementation complexity.

\vspace{-10pt}
\begin{tcolorbox}[colframe=black,colback=gray!10,boxrule=0.8pt,arc=3pt,left=4pt,right=4pt,top=4pt,bottom=4pt]
\textbf{RQ1 Summary:} Encoder-only models achieve 70.4–78.4\% accuracy in ERAP tasks. RoBERTa-large (78.4\%) performs best on text-only input, while ELECTRA-base (78.2\%) excels with profile data. XLNet-large sees the biggest recall boost (+14.4\%) for ``Approved'' when using profiles while DeBERTa-v3-large's accuracy drops to 70.4\%, showing mixed effects of profile data.
\end{tcolorbox}

\subsection{RQ2: Performance of Decoder-only Models}  \label{subsec:decoder}
\begin{table*}[h]
    \caption{Experimental Results on Decoder-only Models}
    \label{tab:my-table}
    \centering
    \resizebox{\textwidth}{!}{
        \begin{tabular}{cccccccc|ccccccc|ccccccc} 
            \hline  
            \multirow{3}{*}{} & \multicolumn{7}{c|}{Text} 
                              & \multicolumn{7}{c|}{Text+Creator Profile} 
                              & \multicolumn{7}{c}{Text+Creator Profile+10 Examples} \\ 
            \cline{2-22}  
                              & \multirow{2}{*}{Acc (\%)} 
                              & \multicolumn{3}{c}{Approved} & \multicolumn{3}{c|}{Rejected} 
                              & \multirow{2}{*}{Acc (\%)} 
                              & \multicolumn{3}{c}{Approved} & \multicolumn{3}{c|}{Rejected}  
                              & \multirow{2}{*}{Acc (\%)}
                              & \multicolumn{3}{c}{Approved} & \multicolumn{3}{c}{Rejected} \\ 
            \cline{3-8} \cline{10-15} \cline{17-22}  
                              &                                     
                              & P (\%) & R (\%) & F (\%) & P (\%) & R (\%) & F (\%)  
                              &                                     
                              & P (\%) & R (\%) & F (\%) & P (\%) & R (\%) & F (\%)  
                              &                                     
                              & P (\%) & R (\%) & F (\%) & P (\%) & R (\%) & F (\%) \\ 
            \hline  
            GPT-3.5-turbo     & 66.4  & 61.9 & 50.4 & 55.6 & 68.7 & 77.9 & 73   
                              & 75.9  & 71.2 & 70.8 & 71   & 79.2 & 79.5 & 79.4
                              & 73.3  & 65.7 & 75   & 70   & 80.2 & 72.1 & 75.9 \\ 
            GPT-4o-mini       & 66.7  & 62.6 & 49.6 & 55.4 & 68.7 & 78.8 & 73.4 
                              & 74.6  & 69.5 & 69.7 & 69.6 & 78.3 & 78.1 & 78.2 
                              & 72.9  & 65.2 & 74.6 & 69.6 & 79.8 & 71.7 & 75.5 \\ 
            DeepSeek-V3       & 61.2   & 53.2 & 57.8 & 55.4 & 67.9 & 63.7 & 65.7 
                              & 74.6  & 67   & 76.7 & 71.6 & 81.5 & 73.1 & 77   
                              & 73.8  & 66.5 & 75.1 & 70.5 & 80.4 & 72.9 & 76.5 \\ 
            Llama 3.1 8B Instruct & 65.7  & 70.6 & 30.4 & 42.5 & 64.7 & \textbf{90.9} & 75.6 
                                  & 78.3   & 75.2 & 71.4 & 73.3 & 80.3 & 83.2 & 81.7 
                                  & 69.5   & 78.6 & 36.8 & 50.1 & 67.3 & \textbf{92.8} & 78 \\ 
            multiple fine-tuned Llama 3.1 8B  
                              & \textbf{76.9} & 75   & 66.9 & 70.7 & 78   & 84   & \textbf{80.9} 
                              & 78.2 & 74.6 & 72.4 & 73.5 & 80.7 & 82.3 & 81.5 
                              & 78.3 & 75.7 & 70.5 & 73   & 79.9 & 83.9 & 81.8 \\  
            binary fine-tuned Llama 3.1 8B  
                              & 75.8 & \textbf{78}   & 58.3 & 66.7 & 74.8 & 88.2 & \textbf{80.9} 
                              & 77.8 & \textbf{77.9} & 65.3 & 71   & 77.8 & \textbf{86.8} & 82 
                              & 78.8 & 77.2 & 69.8 & 73.3 & 79.8 & 85.3 & 82.4 \\ 
            multiple fine-tuned Llama 3.1 8B Instruct
                              & 76.6 & 73   & \textbf{69.4} & \textbf{71.1} & \textbf{78.9} & 81.7 & 80.3
                              & 79   & 74.2 & 76.1 & 75.1 & 82.6 & 81.1 & 81.8 
                              & 78.7 & 73.8 & \textbf{75.7} & \textbf{74.7} & \textbf{82.3} & 80.8 & 81.6 \\  
            binary fine-tuned Llama 3.1 8B Instruct 
                              & 75.5 & 76.2 & 59.9 & 67.1 & 75.2 & 86.6 & 80.5
                              & 77.8 & 76.3 & 67.7 & 71.8 & 78.7 & 85   & 81.7 
                              & 78.6 & 75.9 & 71.2 & 73.5 & 80.3 & 83.9 & 82 \\ 
            Vote
                              & 76.4 & 75.3 & 64.3 & 69.4 & 76.9 & 85.0 & 80.7
                              & \textbf{80.7} & 75.4 & \textbf{79.8} & \textbf{77.5} & \textbf{84.9} & 81.4 & \textbf{83.1}
                              & \textbf{79.7} & \textbf{80.2} & 68.3 & 73.8 & 79.5 & 87.9 & \textbf{83.5} \\ 
            \hline  
        \end{tabular}
    }
    
\end{table*}

In contrast to encoder-only models, decoder-only models initially struggle with the ERAP task. As shown in Table~\ref{tab:my-table}, in zero-shot settings, the models exhibit notably lower accuracy compared to their encoder-based counterparts. However, with the introduction of creator profile and fine-tuning techniques, their performance improves significantly, highlighting the potential of decoder-based architectures when provided with additional learning signals.

The base version of Llama 3.1 8B without fine-tuning was not used; instead, the instruct version was employed. This choice was made because the base model struggled to follow prompts effectively for this task, often failing to generate useful outputs.

When only text was provided, among the tested decoder-only models, GPT-3.5-turbo and GPT-4o-mini achieve moderate performance, with accuracy scores around 66--67\% in the zero-shot setting. DeepSeek-V3 performs the worst initially, with an accuracy of 61.2\%, indicating its difficulty in distinguishing approved and rejected reports and without explicit training data. In contrast, Llama 3.1 8B Instruct, despite showing highly imbalanced class-wise performance in its instruct-tuned variant. LoRA fine-tuned Llama models demonstrate strong stability, achieving high accuracy and consistent performance across all three strategies.

When incorporating creator profile information, all models see improvements across multiple metrics. GPT-based models gain approximately 9 percentage points in accuracy, while DeepSeek-V3 also sees a notable improvement from 61.2\% to 74.6\%. Among all models, Llama 3.1 8B Instruct benefits the most from the inclusion of creator profile information, it significantly improves the recall for approved cases (30.4\% to 71.4\%), contributing to a more balanced classification and a substantial boost in overall accuracy. Similarly, the fine-tuned Llama models achieve a better balance between precision and recall, leading to a more reliable classification outcome. These results show that profile-based context significantly enhances the decoder-only models' ability to differentiate between approval and rejection patterns.  The relatively small performance gap between decoder models of different sizes indicates that adherence to the domain-specific rules inherent in ERAP classification is perhaps less sensitive to general model scale, relying more heavily on how well the prompt conveys these specific task constraints.

However, while the inclusion of creator profile information generally improves prediction accuracy, it also introduces potential biases. In the absence of creator profile information, a total of 184 ER were misclassified by all eight tested decoder-only models, with only 86 cases involving either regular users being incorrectly rejected or inner-developers being incorrectly approved. In contrast, when profile information was included, the number of universally misclassified ER increased to 231, with 218 exhibiting this pattern. This suggests that although profile information enhances model predictions, it may also lead to over-reliance on the creator's role, causing models to be misled in cases where the role information conflicts with the actual merit of the enhancement request.

Introducing 10 few-shot examples failed to further enhance model performance. Notably, GPT-3.5-turbo and GPT-4o-mini experienced slight decreases in accuracy, while DeepSeek-V3 also saw a minor drop. Llama 3.1 8B Instruct, however, exhibited a more significant decline.

Fine-tuning, while only slightly improving the highest accuracy, significantly enhances the stability of decoder-only models across different input settings. LoRA-based fine-tuning helps maintain consistent performance, reducing variability in predictions. Even when only textual information is available, fine-tuned Llama models achieve accuracy levels close to those of models utilizing both text and profile information. This suggests that fine-tuning enables the model to better capture approval patterns, reducing its dependency on additional profile details.
All four fine-tuned models demonstrate strong and consistent performance across different classification strategies. Regardless of whether the task is framed as a binary classification or a more granular multi-category classification, fine-tuning enables the models to achieve a good balance between precision and recall. These results further confirm that fine-tuning significantly enhances the robustness and generalization ability of decoder-only models, allowing them to effectively distinguish between approved and rejected reports.

To further utilize this diversity, we applied a majority voting strategy among the eight models, where an enhancement is classified as approved if more than four models predict it as such. This ensemble approach led to an improvement in accuracy, raising the highest individual model accuracy from 79\% to 80.7\%. These results indicate that aggregating multiple model outputs can help mitigate individual model errors, offering a more robust classification strategy.
Among the 4049 tested ERs, only 231 are wrongly predicted by all eight decoder-only models we employed. This means that at least one model successfully predicted nearly 94.3\% of the enhancements. This observation highlights the strong potential of LLM in ER classification. The diversity in model predictions suggests that different models capture complementary aspects of the data, which could be further leveraged to improve classification performance.

Detecting duplicate ER poses a notable challenge for LLM, as it requires nuanced understanding of previously submitted reports and their similarities. When analyzing the inconsistencies in model predictions under the text + creator profile setting, we observed that a significant number of duplicate reports were misclassified. Without fine-tuning, models often labeled these enhancement reports as ``Approved,'' leading to an average of 192 wrongly accepted duplicates. LoRA fine-tuned models made fewer such mistakes, with the number dropping to around 137 on average.
For example, the enhancement request ``Commenter should be able to edit his last comment'' was a duplicate of a previously submitted request. Among the eight decoder-only models tested, all four without fine-tuning classified it as ``Approved,'' failing to recognize its duplicate nature. In contrast, all four LoRA fine-tuned models correctly classified it as ``Rejected,'' indicating that fine-tuning enhances the model’s ability to identify redundant enhancement requests.

\begin{tcolorbox}[colframe=black,colback=gray!10,boxrule=0.8pt,arc=3pt,left=4pt,right=4pt,top=4pt,bottom=4pt]
\textbf{RQ2 Summary:} Decoder-only models initially underperform but improve with profile information and fine-tuning (79\% accuracy). Few-shot examples do not enhance performance. Profile-based context boosts accuracy but may introduce biases. LoRA fine-tuning stabilizes predictions, and a voting-based ensemble enhances robustness. However, duplicate detection remains a challenge despite fine-tuning improvements.
\end{tcolorbox}

\subsection{RQ3: Comparison Between LLM and Traditional Methods}

\begin{table*}[h]
    \caption{Experimental Results on Traditional Methods}
    \label{tab:table5}
    \centering
        \begin{tabular}{cccccccc|ccccccc}
            \hline
            \multirow{3}{*}{} & \multicolumn{7}{c|}{Text} & \multicolumn{7}{c}{Text+Creator Profile} \\ \cline{2-15}
            & \multirow{2}{*}{Acc (\%)}  & \multicolumn{3}{c}{Approved} & \multicolumn{3}{c|}{Rejected} & \multirow{2}{*}{Acc (\%)}  & \multicolumn{3}{c}{Approved} & \multicolumn{3}{c}{Rejected} \\ \cline{3-8} \cline{10-15}
            & & P (\%) & R (\%) & F (\%) & P (\%) & R (\%) & F (\%) & & P (\%) & R (\%) & F (\%) & P (\%) & R (\%) & F (\%) \\ \hline
            LSTM-GLOVE & \textbf{74}   & 75.1 & \textbf{56.4} & \textbf{64.4} & \textbf{73.5} & 86.6 & \textbf{79.5}  
                       & 67.6 & 60.5 & \textbf{64.1} & \textbf{62.3} & \textbf{73.2} & 70.1 & 71.6 \\
            LSTM-BERT & 58.3 & 0    & 0    & 0    & 58.3 & \textbf{100}  & 73.7  
                      & \textbf{69.1} & \textbf{81.5} & 33.3 & 47.2 & 66.5 & 94.6 & \textbf{78.1} \\
            CNN-GLOVE & 68   & \textbf{79.8} & 30.9 & 44.6 & 65.7 & 94.4 & 77.5  
                      & 67.9 & 80.3 & 30.4 & 44.1 & 65.6 & \textbf{94.7} & 77.5 \\
            CNN-BERT & 58.3 & 0    & 0    & 0    & 58.3 & \textbf{100}  & 73.7 
                     & \textbf{69.1} & 75.6 & 38   & 50.6 & 67.3 & 91.2 & 77.5 \\
                     \hline

                     \hline
        \end{tabular}

\end{table*}
Traditional methods, particularly CNN-BERT from prior studies, achieved 83.6\% accuracy in non-cross-application validation and 80.7\% in cross-application validation when evaluated through 10-fold cross validation~\cite{niu2023utilizing}. However, these results were obtained under experimental conditions that allowed potential data leakage between training and test sets. When we strictly reimplemented them using chronological data splitting (i.e., training on earlier data and testing on subsequent data to simulate real-world deployment), their performance dropped significantly to approximately 70\% accuracy. This decrease is primarily due to changes in class distribution rather than an inherent loss of model capability—under chronological splitting, the proportion of Approved reports is higher, making the class distribution more balanced and leading to different overall accuracy dynamics. The detailed reproduction results are provided in Table~\ref{tab:table5}.

As shown in Table~\ref{tab:table3}, modern encoder-only LLM achieve comparable accuracy (70.4-78.4\%) to traditional methods under our rigorous evaluation protocol, while demonstrating better class balance. For instance, DeBERTa-v3-base attains F1-scores of 70.9\% (Approved) and 81.9\% (Rejected) when incorporating profile information, compared to CNN-BERT's 58.8\% and 89.8\% in the original study~\cite{niu2023utilizing}. This indicates that while traditional methods excel at identifying majority-class samples (Rejected), LLM better handle the imbalanced nature of the task through contextual understanding.

Two key advantages of LLM emerge from our analysis:
\begin{enumerate}
\item \textbf{Class-wise Balance}: The F1-score ratio between Approved and Rejected classes improves from approximately 1:1.5 (CNN-BERT) to 1:1.1 (fine-tuned Llama 3.1 8B Instruct), critical for practical deployment where both error types carry significant costs.

\item \textbf{Feature Learning Capacity}: Without manual feature engineering, LLM automatically capture subtle approval signals like request specificity (e.g., ``show confidential bugs using color'' vs. vague ``Improve security'') and creator expertise level.
\end{enumerate}

Notably, traditional methods show weakness in handling approved reports (F1 $\leq$ 60.9\% in the original study), often misclassifying them as rejected due to their minority status~\cite{niu2023utilizing}. In contrast, LLM like XLNet-large-cased with profile information achieve 72.2\% F1 for approved reports through contextual understanding of creator profile - a capability absent in CNN-BERT's bag-of-words approach.

Our experiments suggest that the high accuracy of traditional methods reported in prior studies may be influenced by cross-validation methodologies that do not fully consider temporal dependencies in real-world ERAP systems. When re-evaluated using a chronological data split, we find that modern LLM achieve comparable or superior performance while providing better interpretability through attention patterns that align with human decision-making factors. 

A critical insight from our findings is the importance of recall for \textit{Approved} reports in the ERAP. In practical deployment, failing to identify an enhancement request that should be approved can delay meaningful software improvements, making recall for \textit{ Approved} cases a crucial metric. Traditional methods, such as CNN-BERT, tend to struggle with \textit{Approved} reports due to their minority status in the dataset, often misclassifying them as \textit{Rejected}. In contrast, fine-tuned LLM significantly improve \textit{Approved} recall while maintaining balanced classification performance, addressing this limitation effectively.

To further understand the limitations of LLM, we analyzed 126 ER that were consistently misclassified across all 18 tested LLM. 70.63\% of these misclassified ER involve technical aspects related to API compatibility, performance optimization, and UI/UX improvements. A common characteristic of these ER is the presence of domain-specific terminology and highly technical descriptions. To explore the reasons behind these misclassifications, we selected representative errors and re-predicted them using decoder-only models. We then provided the process of predictions as context, prompting them to generate specific explanations for why the ER were rejected.

\begin{figure}[h]
    \centering
    \includegraphics[width=1\linewidth]{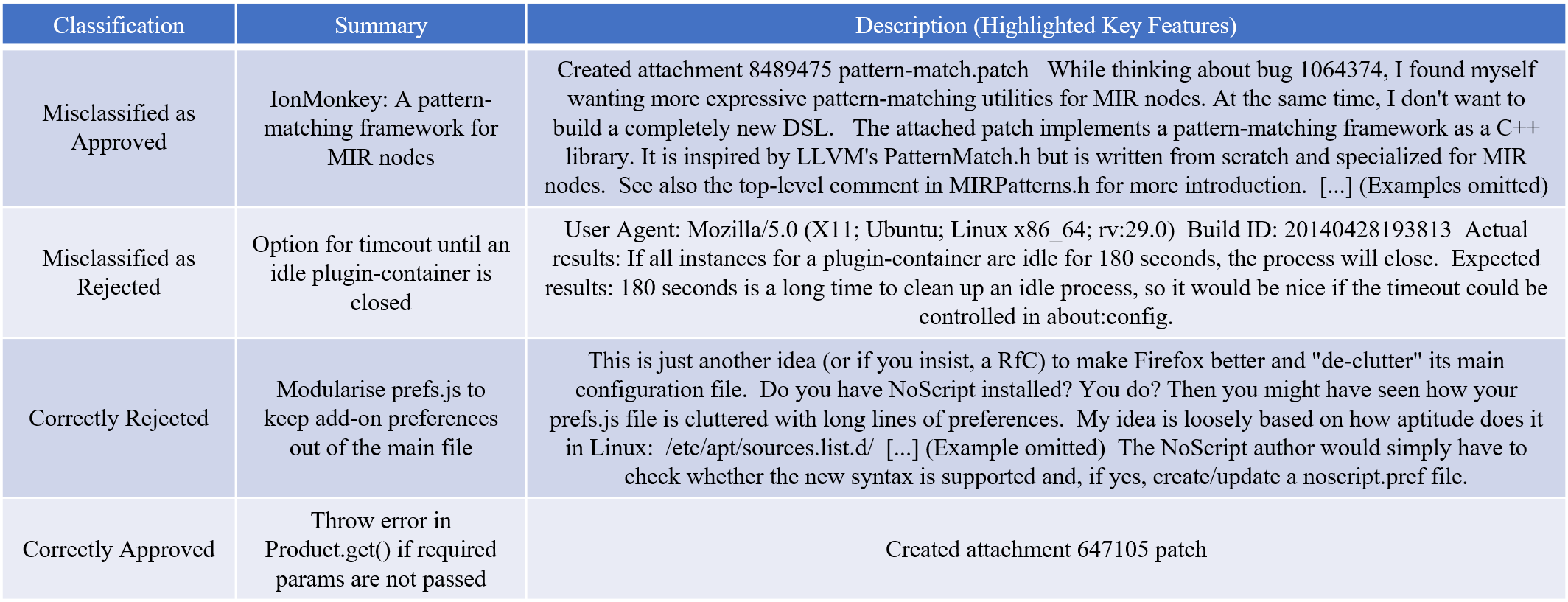} 
    \caption{Examples}
    \label{fig:examples}
\end{figure}

Our analysis suggests that the issue is not that LLM fail to understand technical concepts but rather that they struggle to assess whether a proposed technical improvement aligns with the product’s needs. When faced with well-articulated enhancement requests that include concrete implementation details, LLM are more likely to be persuaded into approving them, even if they may not be necessary for the product. For example, in the case of the enhancement request ``IonMonkey: A pattern-matching framework for MIR nodes'', the description presented a highly detailed solution, including a working patch, example usage, and references to similar frameworks.
the model inferred that the proposed change aligned with the project's goals, had high community acceptance, and already had a preliminary implementation (with an attached patch) that had undergone some testing. Consequently, the model incorrectly classified it as an approved enhancement.

Conversely, enhancement requests with less clear descriptions or those lacking specific implementation details were more likely to be rejected by the LLM. For instance, the request ``Option for timeout until an idle plugin-container is closed'' was misclassified as rejected because the model perceived its impact as low-priority and was concerned about potential downsides such as performance degradation, memory leaks, or user misconfigurations. Additionally, since this request originated from a regular user rather than a developer, the model was more inclined to dismiss it.

However, our analysis also revealed cases where non-technical enhancement requests were correctly rejected despite having detailed descriptions. A notable example is ``Modularise prefs.js to keep add-on preferences out of the main file'', which proposed restructuring Firefox’s configuration file by modularizing preferences into separate files. While the request was well-articulated and supported by a concrete analogy (Linux’s sources.list.d structure), it lacked a compelling justification for why such a change was necessary. The model correctly recognized that the proposal, despite its clarity and thorough explanation, did not address a critical issue and therefore classified it as rejected.

On the other hand, some enhancement requests with minimal descriptions were still correctly accepted. For example, the request ``Throw error in Product.get() if required params are not passed'' contained only a brief summary and an attached patch. Despite the lack of detailed reasoning, the model correctly inferred its necessity based on the clear indication that the enhancement addressed a functional correctness issue.

This analysis shows that by enhancing LLM' ability to link domain knowledge with product-specific requirements and ability to differentiate between well-documented but non-essential changes and crucial product-driven enhancements, we can further refine its classification accuracy in real-world enhancement approval tasks.

\begin{tcolorbox}[colframe=black,colback=gray!10,boxrule=0.8pt,arc=3pt,left=4pt,right=4pt,top=4pt,bottom=4pt]
\textbf{RQ3 Summary:} While traditional models struggle with \textit{Approved} reports, LLM significantly enhance their recall by 12\%, reducing the Approved-to-Rejected F1-score ratio from about 1:1.5 (CNN-BERT) to about 1:1.1 (fine-tuned Llama 3.1 8B Instruct). This suggest that LLM are well-suited for deployment in real-world ERAP systems, where ensuring fair classification and minimizing false rejections are essential.
\end{tcolorbox}

\section{Threats to Validity}\label{sec:threats}

One potential threat to validity in our study is the limitation in model selection due to cost constraints. While we have employed a diverse set of decoder-only models, we did not include some of the most advanced models available, such as Claude 3.7 Sonnet, GPT-4.5, OpenAi o1 and DeepSeek-R1. As a result, our findings may not fully reflect the state-of-the-art performance achievable with the latest LLM.

Another limitation arises from the dataset's temporal constraints. The data used in our experiments spans from 1997 to 2016, which may have been included in the training data of the evaluated models. Consequently, when applied to repositories created after the models' training cutoff dates, their performance might not be as strong due to a lack of exposure to more recent trends and changes in ER characteristics.

\section{Conclusion and Future Work}\label{sec:conclusion}

Our study demonstrates the effectiveness of LLM in the ERAP task. Among encoder-only models, RoBERTa-large achieves the best performance with 78.4\% accuracy, exhibiting strong classification capabilities and stable predictions. For decoder-only models, fine-tuned LLaMA 3.1 8B Instruct performs the best, reaching 79\% accuracy. While decoder-only models initially struggle, they improve significantly with profile context and fine-tuning.

When we evaluated models on the time-ordered dataset, traditional methods exhibited a decline in accuracy over time, while LLM maintained a strong performance advantage, achieving 5\% higher accuracy and a 12\% higher APPROVED recall compared to traditional approaches.

Analyzing inconsistent predictions and incorrect ER, we find that fine-tuned decoder models excel at detecting duplicates. However, incorporating creator profile information can sometimes lead decoder models to overly rely on creator role. Additionally, the presence of domain-specific terminology and highly technical descriptions makes LLM more susceptible to well-articulated enhancement requests that include concrete implementation details, increasing the risk of being misled by technically detailed yet flawed ERs.

Fine-tuning stabilizes model predictions, reducing dependency on additional inputs. Despite challenges in handling technical descriptions and duplicate detection, fine-tuned LLM offer a viable alternative to traditional methods. Their flexibility and improved recall make them well-suited for real-world ERAP applications.

As highlighted in Section~\ref{subsec:decoder}, at least one model correctly predicts 94.3\% of the data, demonstrating the high upper bound of decoder-only models in ER classification. This suggests that while individual models may struggle with certain cases, the collective knowledge across multiple models captures a more comprehensive understanding of the data. However, discrepancies in model predictions indicate that there is still room for improvement, particularly in handling ambiguous or borderline cases.

A promising research direction is to explore methods that capitalize on the diversity of model predictions. For ER with inconsistent results, ensemble strategies can be applied, possibly incorporating weighting schemes based on each model’s past accuracy. Alternatively, a specialized AI agent could serve as the final arbiter, integrating auxiliary information to make informed decisions. Encoder-based models could provide explanations and reasoning traces for their predictions, which may serve as additional input for the agent. Furthermore, techniques like chain-of-thought prompting could guide decoder-only models to reassess their outputs, enhancing stability and reducing misclassification.

Enriching the input context is another viable approach to improving model performance. Integrating external sources—such as release notes, product documentation, and relevant technical materials—into a retrieval-augmented generation (RAG) system could help models gain deeper insights into the evolution and current state of the product. This would not only support more accurate alignment with product improvement goals but also improve duplicate detection and reduce vulnerability to misleading but well-written ER. Future work may also explore hybrid approaches, combining LLM with traditional classifiers to leverage complementary strengths and further enhance ERAP performance.

\section{Data Availability}
The code and dataset used in this study are publicly available at \url{https://github.com/ZUOHS/enhancement}
\bibliographystyle{ACM-Reference-Format}
\bibliography{sample-base}










\end{document}